# The main role of fractal-like nature of conformational space in subdiffusion in protein

*Luca Maggi*

*Mechanisms of Diseases, Institute for Research in Biomedicine (IRB Barcelona), The Barcelona Institute of Science and Technology, Baldiri Reixac 10-12, Barcelona 08028, Spain*

*e-mail address: luca.maggi@irbbarcelona.org*

## Abstract:

Protein dynamics is a fundamental element to comprehend their biological functions. However, a theoretical picture providing microscopic-detail explanation of its relevant features is still missing. One of the outmost relevant properties exhibited by this dynamic is its subdiffusivity, whose origins are still unknown. Here, by directly comparing all-atom molecular dynamics simulations and theory we show that this behavior mainly arises from the fractal nature of the network of metastable state of conformational state over which protein dynamics, thought as diffusion process, takes place. This process is assumed to be Markovian by the employed theoretical picture. Therefore, to further support its validity, we built a simple Markov state model starting from the simulations outcome and show that it exhibits a subdiffusive behavior, in quantitative agreement with the one associated to the molecular dynamics. Moreover, Molecular dynamics gives direct access to relevant quantities which allowed us to rule out the possibility the Continuous Time Random Walk can explain the protein subdiffusivity.

## Main text:

A fundamental paradigm in structural biology stated the relationship between protein structure and function. However, aside a single specific arrangement of protein atoms (i.e., a conformation), the dynamics is widely being recognized as a pivotal element to understand protein function at microscopic level [1,2]. Catalytic enzymatic reactions [3], signal transduction [4] and molecule transport across the plasmatic membrane [5] are significant examples in which protein dynamics plays a crucial role and, thus, its investigation cannot thus be neglected. To avoid any source of confusion here we refer to protein dynamics as every time-dependent change of protein conformation. This is determined by the interactions that residues have with each other and with external environment. The whole set of those produces a potential energy landscape whose explicit analytical treatment is, however, unfeasible mainly due to its high dimensionality and complexity. Nevertheless, experimental and computational studies highlighted some of its relevant features as the inherently "roughness" which gives rise to a conformational space composed by a plenty of metastable states separated by energy barriers of different heights [ [1,2,6]. According to this picture, therefore, protein dynamics is often modeled as a diffusion process among those metastable states [7–9], the same model is adopted in the present work. Previous investigations directly and indirectly showed [8,10–14] that the diffusion process exhibits a subdiffusive behavior, which implies a sublinear relationship between the mean square displacement (MSD) and time. More formally speaking, if $X(t) = \{x_1(t), \ldots x_N(t)\}$ is a single conformation, with $x_i(t)$ the i-th degree of freedom at time t and N the total number of degrees of freedom, it follows:

$$MSD = \langle |X(0) - X(t)|^2 \rangle \sim t^\alpha \qquad (\text{Eq. 1})$$

Where <..> represents an ensemble average. In Eq. 1 subdiffusivity imposes α to be smaller than 1. Despite numerous and notably theoretical efforts [14–17] the microscopic origin of this phenomenon is still unclear. In this letter we bring strong evidences that indicates the fractal nature of the protein conformational space as the main origin of subdiffusion in protein. This is achieved by employing all atom Molecular Dynamics (MD) simulations to directly verify theoretical results. The theory which describes a diffusion process on a fractal structure prescribes that exponent α is given by [18,19]:

$$\alpha = \frac{d_s}{d_f} \qquad (\text{Eq. 2})$$

Where $d_f$ is associated to geometrical properties of conformational space and $d_s$ is related to spectral features of the operator generating the diffusion process [18,19]. Our goal is, therefore, to calculate α, $d_f$, and $d_s$ separately and verify whether Eq. 2 holds. The evaluation of those three quantities has been achieved by the analysis of 1-us MD simulations of three biomolecules: the N-terminal of the human histone H4 tail (H4); The Villin headpiece (Villin; PDB ID: 1VII [20]) and a PDZ domain (PDZ; PDB ID: 1D5G [21]), Fig. 1a-c. They differ for the number of residues, which are: 25, 32 and 96 for H4, Villin and PDZ respectively, and its secondary and tertiary structure as H4 is a totally disordered peptide, Villin exhibits a partial structure and the PDZ can be classified as a small globular structured protein. All the MD simulations details are given in SI. In order to reduce the number of degrees of freedom we projected all the trajectories on to the first two principal component extracted from a principal component analysis, PC1(t) and PC2(t), carried out taking into account Cα's backbone only. Hence, those two variables define a single sampled conformation, X(t)={PC1(t), PC2(t)}. The identification of metastable states has been done over this sub-space by means of a clustering algorithm. In this work an agglomerative hierarchical clustering method is employed [22]. The reasons leading to this choice are two-fold. Firstly, the intrinsic hierarchy of the whole metastable states appears reasonable as structural differences among conformations can be naturally classified as sub-sets of decreasing size which sub-divide the entire space. This idea is supported by previous works which highlighted this feature [2,7]. On the other hand, this clustering method presents technical advantages as it does not require to set a fixed number of clusters (as K-means methods) employing an adjustable parameter that controls the cluster average size ($\varepsilon$) and it does not produce any outlier conformations which are hard to be included in the theoretical picture. The dynamics including each single conformation is, thus, replaced by a coarse-grained one involving only the cluster centroids which correspond to the representative conformations of each single metastable state. The average cluster size, is chosen to reproduce at the best the MSD calculated from fine-grained conformational sub-space, still providing a reliable sampling of each cluster (see Fig SI 1). The MSD calculation are performed using a moving average to cancel out the dependence from the initial conditions, and it reads:

$$MSD = \frac{1}{T-t}\int_0^{T-t} d\tau \ |X(t+\tau) - X(\tau)|^2 \qquad \text{(Eq. 3)}$$

Where t < 0.01 T, where T is maximum simulation time (1us). We found that setting $\varepsilon$ within a range from 1.0 to 0.2, depending on the system, produces almost the same subdiffusive behavior as showed by the very small relative difference between the exponent α's evaluated in the two cases, (reported as percentage) $\Delta = \frac{|\alpha - \alpha_{fine}|}{\alpha_{fine}}$, where α and $\alpha_{fine}$ are the exponents calculated for coarse- and fine-grained conformational space respectively (Fig 1d–f). The distribution of metastable states over the sub-space is connected to $d_f$ which relates the number of cluster centroids within a sphere of radius r, M(r), to the radius itself as, $M(r) \sim r^{d_f}$. The M(r) profiles are evaluated averaging over all the clusters and presented in Fig. 2a-c. All of them exhibit a power law relation with $d_f$ always less than 2 showing the non-homogenous distribution of cluster centroids. While $d_f$ is associated with the geometric arrangement of metastable states in the conformational sub-space, $d_s$ is related to their "connectivity", determining the spectral density of the operator generating the diffusion process [23]. Employing an operative definition this exponent characterizes the probability of a trajectory to return to its starting point ($P_o$) after a time t, being $P_o \sim t^{-d_s/2}$. In our case the "starting point" coincides with the starting metastable states (i.e., starting cluster). Hence, we introduced $C(t+\tau, \tau)$ which is a function equal to 1 if the cluster visited at $t+\tau$ and $\tau$ are the same and 0 otherwise and Po is calculated employing a moving average:

$$P_o(t) = \frac{1}{T-t}\int_0^{T-t} d\tau \ C(t+\tau, \tau) \qquad \text{(Eq. 4)}$$

The profiles show a good agreement with power-law relation within about four order of magnitude (Fig. 2 d-f). In Tab.1 we summarize all the calculated quantities and compare the subdiffusion exponent α with $d_s/d_f$. We found an excellent quantitative agreement between these two. Therefore, it turns out that the theory of diffusion on fractals is capable to adequately modeling the protein conformational space exploration as described by all atom Molecular dynamics simulations and it should be noticed the general validity of this finding as it holds for biomolecules which present very different structural features. This is the main result of this work.

Interestingly, the correspondence between theory and simulations entails an important feature of diffusion process which is its Markovianity [19]. Therefore, to test this implication we have extracted a transition probability matrix (T) directly form the MD simulations and analyzed the MSD resulting from this Markov state model (MSM). The matrix T is $N_c$ x $N_c$ matrix, where $N_c$ is the total number of clusters, and each element Tij is equal to:

$$T_{ij} = \frac{s_{ij}}{\sum_{i=1}^{N_c} s_{ij}} \qquad \text{(Eq. 5)}$$

Where $s_{ij}$ is the number of "jumps" between the I-th and j-th cluster. Obviously, the finiteness and discreteness of T prevents the results to arbitrarily coincide with the MD results which are, instead, produced by an infinite and continuous operator [24]. To each cluster has been assigned a stationary probability value ($\boldsymbol{\pi}_s$) extracted directly from the simulations and system time evolution consist in the evolution of the Nc-length probability vector $\boldsymbol{\pi}$, which evolves like a time discrete Markov chain and at each step n we can calculate this vector for the step n+1 as:

$$\boldsymbol{\pi}_{n+1} = T \, \boldsymbol{\pi}_n \tag{Eq. 6}$$

Therefore, the MSD in this case has been calculated as:

$$MSD = \Delta t \cdot \sum_{i=1,j=1}^{N_c} \pi_s^i \cdot \pi_n^j \, d_{ij}^2 \tag{Eq. 7}$$

Where Δt is the smallest sampling step in MD simulations (20 ps, see SI). The apices I,j indicates the relative cluster and $d_{ij}^2$ is the square distance between them. Despite the unavoidable flaws introduced in the presented MSM, the calculated MSD exhibits a very similar quantitative time dependence to the one associated to MD simulations (Fig 3). This supports further the main finding and gives a new impetus to the debate over the long-memory effect on protein dynamics which has been previously accounted for to model the diffusion process in the conformational space. One of the most common microscopic theoretical pictures involving memory effects is the continuous time random walk (CTRW) which describes the diffusion as time-continuous jumping process among metastable states separated by energy barriers which, to give rise a subdiffusive dynamics, are distributed according to a power-law. This produces a sub-linear time relationship for the average number of jumps $\langle N(t) \rangle \sim t^\alpha$ which originates subdiffusion [25]. To show that the subdiffusivity shown in all atom MD simulations cannot be described by CTRW we have directly calculated <N(t)> as:

$$\langle N(t) \rangle = \int_0^t d\tau \, |C(\tau + \Delta t, \tau) - 1| \tag{Eq. 8}$$

This quantity shows a time linear dependence independently from the size of clusters (Fig 4a). Moreover, the probability of "observing a jump within a time t, presents an exponential decay since its profile doesn't show any region following a power law relationship which should be invariant upon cluster size changing (Fig 4b). Therefore, the evaluation of those two quantities rules out CTRW as a possible description of subdiffusivity observed in all atom MD simulations.

In conclusion we brought compelling evidence that subdiffusive protein dynamics, as described by MD simulations, stems from the fractal nature of the conformational space. The high dimensional and rough potential energy landscape gives rise to separated basins of attractions, namely metastable states whose distribution in the conformational space, regulated by $d_f$, resembles a fractal structures. On the other hand, $d_s$ is related to the connections among those states modulating the accessibility from one to another. The whole dynamics is therefore described by these two exponents which are directly connected to the geometry of the conformational energy landscape. As corollary we also highlighted the parallelism between the diffusion and a Markov process as implied by theory of diffusion on fractals and ruled out the role of CTRW in the origin of subdiffusion phenomenon.

**Figures:**

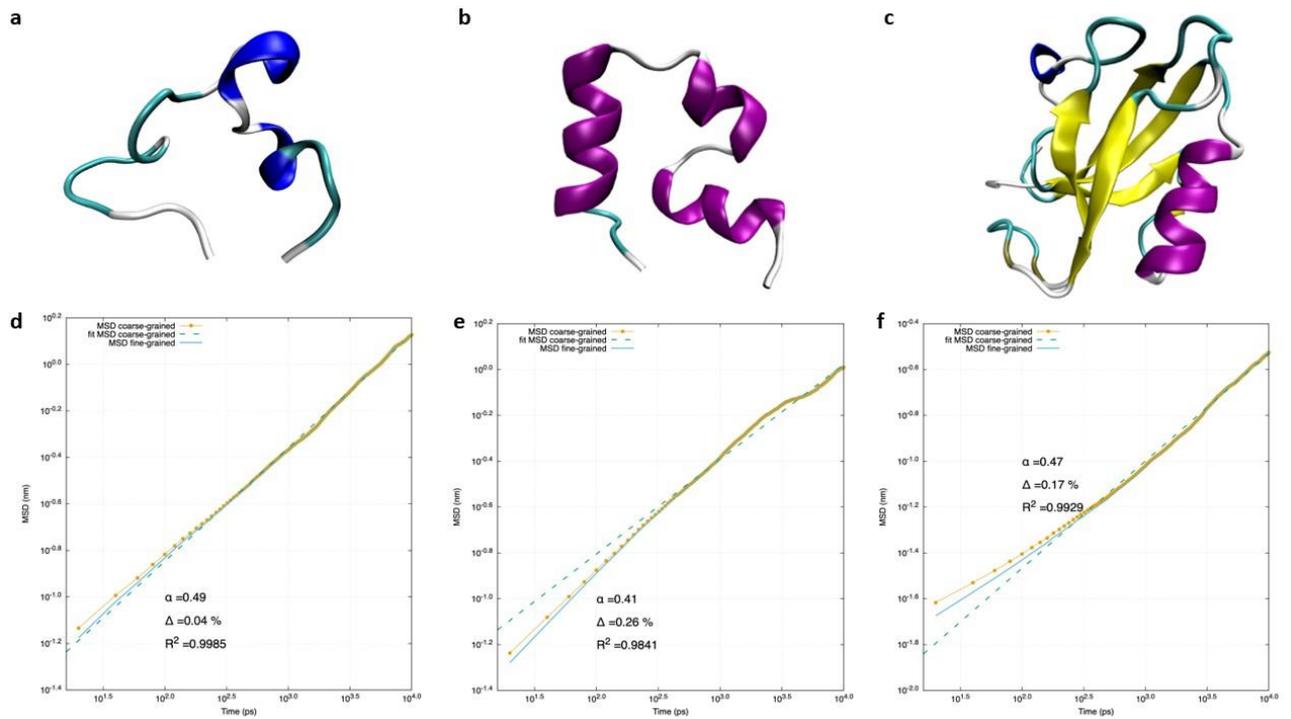

**Fig 1:** Snapshot extracted from MD simulations of the three investigated systems and the associated MSD linearly fitted: a-d) H4 tail, b-e) Villin, c-f) PDZ domain and

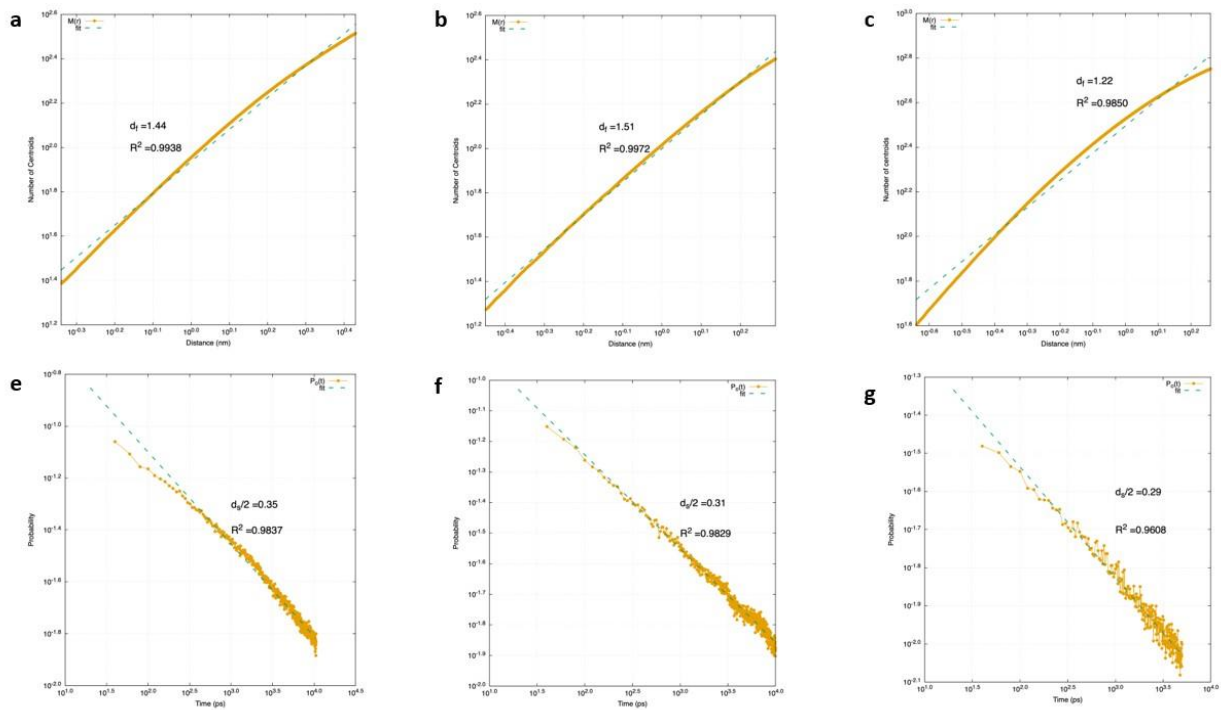

**Fig 2:** M(r) and $P_o(t)$ profile and the associated linear fit for the three systems: a-d) H4 tail, b-e) Villin, c-f) PDZ domain. The radius value range within which the M(r) linear fit has been performed,

corresponds to 5% and 70 % of the total number of metastable states, to avoid artifacts due to the conformational space discreetness and finiteness

| System | $d_f$ | $d_s$ | $\alpha = d_s/d_f$ | $\alpha_{fit}$ |
|---|---|---|---|---|
| H4 | 1.44 | 0.70 | 0.486 | 0.49 |
| Villin | 1.51 | 0.62 | 0.411 | 0.41 |
| PDZ | 1.22 | 0.58 | 0.475 | 0.47 |

**Tab 1:** Summarizing table showing all the evaluated exponents for all the system under investigation and comparing the α coming from the theory and extracted directly from the fit with MSD ($\alpha_{fit}$).

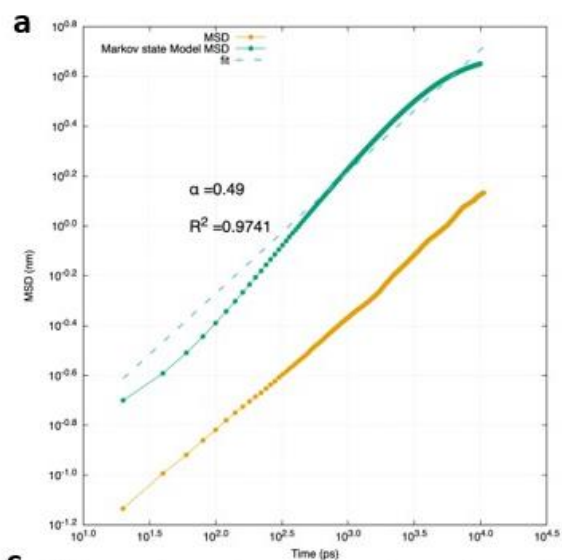

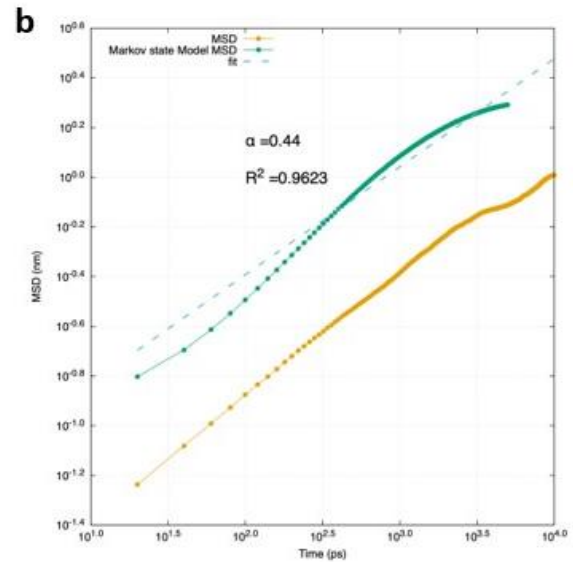

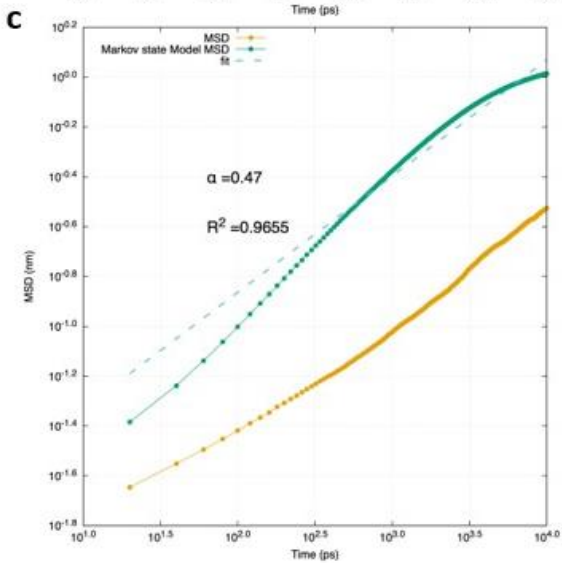

| System | $\alpha_{fit}$ | $\alpha_{MSM}$ |
|---|---|---|
| H4 | 0.49 | 0.49 |
| Villin | 0.41 | 0.44 |
| PDZ | 0.48 | 0.47 |

**Fig. 3:** Comparison between Markov state model MSD and the one calculated form MD simulations for the Different systems investigated: a) H4 tail, b) Villin, c) PDZ d) Summarizing table

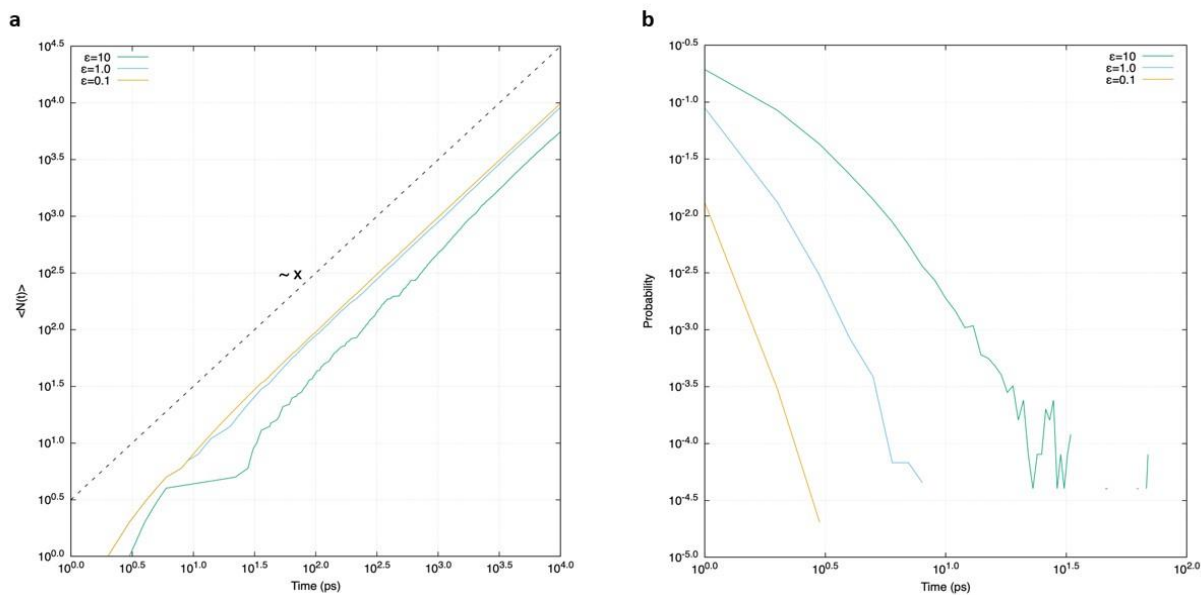

**Fig. 4:** a) <N(t)> and b) probability of jumps for Villin with three different average cluster size. The linear profile for <N(t)> as well as the exponential decay for probability of jump are conserved feature upon changing the cluster size. H4 and PDZ plots exhibit the same characteristics (Fig. SI 2).

## Supporting Information:

*Molecular Dynamics details*

All the presented simulations are carried out with following procedure. The simulations boxes which presented different sizes: 40x40x40, 47x47x47, 69x69x69 $A^3$ for H4, Villin and PDZ respectively are filled with TIP3P water molecules and Na+ and Cl- atoms to neutralize the systems. A first 100 ns equilibration step has been run followed by 1-us production run from which has been extracted frames every 20 ps. LINCS [26] was used to constrain all the bonds involving hydrogens allowing us to employ a 2 fs step to integrate the Newton equations. The first equilibration part is divided into 50 ns NVT ensemble simulation where T=310K, followed by a 50 ns NPT ensemble run to set the total pressure to 1 atm. We employed a Berendsen thermostat [27] {Ref.} with a coupling constant of 0.4 ps for the NVT ensemble and a Nose-Hover [28] {ref.} thermostat along with a Parrinello-Rahman Barostat [29] with 0.4 and 0.6 ps coupling constant respectively for the NPT ensemble, which are used also for the production run. We employed Mesh Ewald method to account for long-range interactions with a real-space cut-off of 12 A. We used GROMAC 2020.2 [30] code with Amber99sb-ildn force field for all the simulation.

*Molecular Dynamics Analysis:*

All the linear fit presented in this work has been done with Gnuplot 5.4. Pyhton Scikit Learn suite [31] is employed for the hierarchical clustering using a Word linkage method. The rest of the analysis are carried out utilizing *ad hoc* in-house scripts.

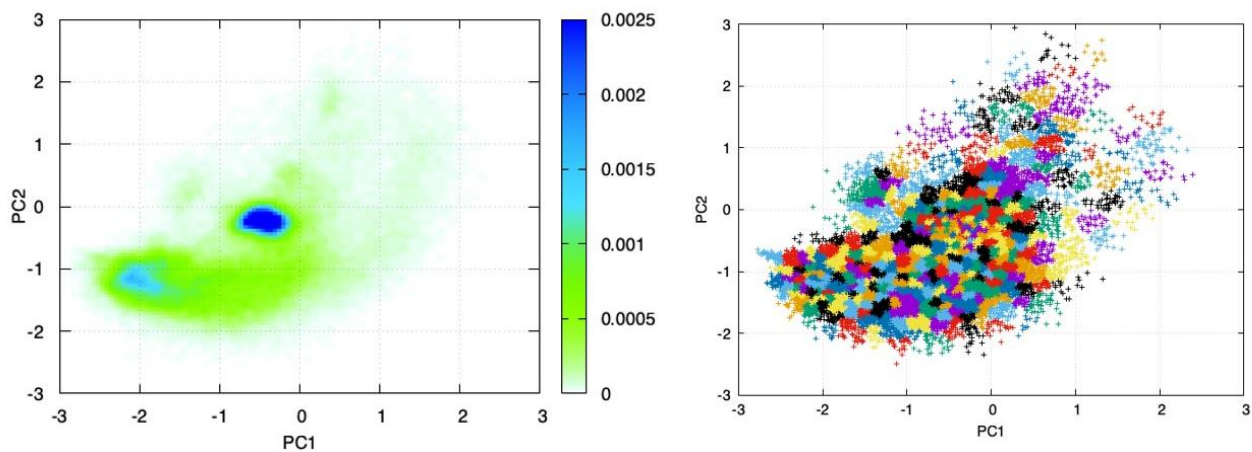

**Fig. SI 1:** Villin free energy surface projected on to PC1 and PC2 on left. On right the same surface is subdivided into clusters identified by hierarchical clustering as described in the main text

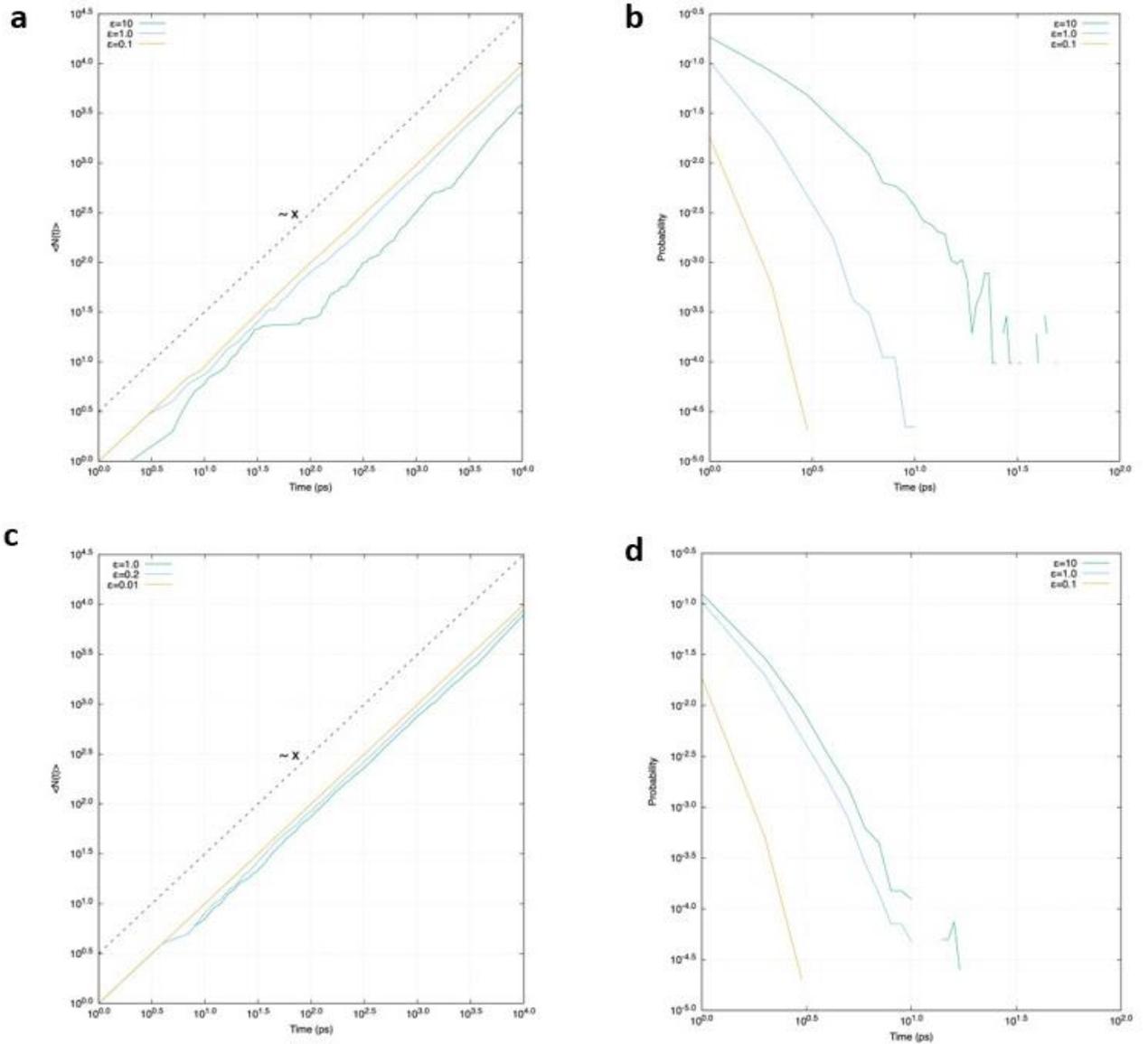

**Fig. SI 2:** <N(t)> and probability of jumps for H4 ( a)-b) ) and PDZ ( c) - d) ).